\documentclass[preprint,number,3p]{elsarticle}

\usepackage{amssymb}
\usepackage{bm}
\usepackage{mathrsfs}
\usepackage{subfigure}
\usepackage{graphicx}
\journal{Physics Letters B}
\begin{document}

\begin{flushright}
\pprinttitle{CYCU-HEP-12-02, HUPD1103}
\end{flushright}

\begin{frontmatter}

%% Title, authors and addresses

%% use the tnoteref command within \title for footnotes;
%% use the tnotetext command for the associated footnote;
%% use the fnref command within \author or \address for footnotes;
%% use the fntext command for the associated footnote;
%% use the corref command within \author for corresponding author footnotes;
%% use the cortext command for the associated footnote;
%% use the ead command for the email address,
%% and the form \ead[url] for the home page:
%%
%% \title{Title\tnoteref{label1}}
%% \tnotetext[label1]{}
%% \author{Name\corref{cor1}\fnref{label2}}
%% \ead{email address}
%% \ead[url]{home page}
%% \fntext[label2]{}
%% \cortext[cor1]{}
%% \address{Address\fnref{label3}}
%% \fntext[label3]{}

\title{
Phase diagram of Nambu--Jona-Lasinio model\\
with dimensional regularization
}

%% use optional labels to link authors explicitly to addresses:
%% \author[label1,label2]{<author name>}
%% \address[label1]{<address>}
%% \address[label2]{<address>}

\author[label1]{T. Inagaki}
\author[label2]{D. Kimura}
\author[label3]{H. Kohyama}
\author[label4]{A. Kvinikhidze}

\address[label1]{
Information Media Center, Hiroshima University,
Higashi-Hiroshima, Hiroshima 739-8521, Japan
}
\address[label2]{
Faculty of Education, Hiroshima University,
Higashi-Hiroshima, Hiroshima 739-8524, Japan
}
\address[label3]{
Department of Physics, Chung-Yuan Christian University,
Chung-Li 32023, Taiwan
}
\address[label4]{
A. Razmadze Mathematical Institute, Tbilisi State University,
}

%%%%%%%%%%%%%%%%%%%%%%%%%%%%%%%%%%%%%%%%%%%%%%%%%%%%%%%%%%%%%%%%%%%%%%
\begin{abstract}
We investigate the phase diagram on temperature-chemical potential
plane in the Nambu--Jona-Lasinio model with the dimensional
regularization. While the structure of the resulting diagram
shows resemblance to the one in the frequently used cutoff
regularization, some results of our study indicate striking
difference between these regularizations. The
diagram in the dimensional regularization exhibits strong
tendency of the first order phase transition.
\end{abstract}
%%%%%%%%%%%%%%%%%%%%%%%%%%%%%%%%%%%%%%%%%%%%%%%%%%%%%%%%%%%%%%%%%%%%%%

\begin{keyword}
QCD phase diagram, Chiral symmetry, Effective field theory
%% keywords here, in the form: keyword \sep keyword
%% MSC codes here, in the form: \MSC code \sep code
%% or \MSC[2008] code \sep code (2000 is the default)
\end{keyword}
%HUPD1103
\end{frontmatter}

%\linenumbers

%% main text
%%%%%%%%%%%%%%%%%%%%%%%%%%%%%%%%%%%%%%%%%%%%%%%%%%%%%%%%%%%%%%%%%%%%%%
%%%%%%%%%%%%%%%%%%%%%%%%%%%%%%%%%%%%%%%%%%%%%%%%%%%%%%%%%%%%%%%%%%%%%%
%    	Sec. 1 INTRODUCTION                                          %
%%%%%%%%%%%%%%%%%%%%%%%%%%%%%%%%%%%%%%%%%%%%%%%%%%%%%%%%%%%%%%%%%%%%%%
%%%%%%%%%%%%%%%%%%%%%%%%%%%%%%%%%%%%%%%%%%%%%%%%%%%%%%%%%%%%%%%%%%%%%%
\section{INTRODUCTION}
\label{intro}
The phase diagram of the quark matter
%s 
has been actively investigated
for decades~\cite{Fukushima:2010bq}. Quarks are confined inside hadrons
and can not be observed as free particles at low energy. On the other
hand at high energy, quarks become free particles due to the asymptotic
freedom of the strong interaction. Therefore, it is expected that
quarks undergo the phase transition between confined and deconfined
states
%. Thus the study of the transition of quark matters is an
%interesting and important issue 
%<<
which is one of the most important issues
%>>
in the theoretical and experimental particle physics.

%The first principle theory for 
The fundamental theory to describe quark matter is quantum
chromodynamics (QCD), the theory of strong interaction.
%whose ultimate goal is to describe all the
%phenomena relating to the strong interaction at entire energy. 
It is, however, %difficult 
not practical to extract reliable predictions at low energy
due to the necessity of complicated nonperturbative calculations
in this area. % nature of quark matters.  Then we need some
For this reason some effective approaches
%to study a quark system at low energy,
are used such as
the Nambu--Jona-Lasinio (NJL) model~\cite{NJL} and its Polyakov-loop
incorporated version, the PNJL model~\cite{Fukushima:2003fw}, the
linear sigma model~\cite{LsM}, 
the chiral perturbation theory~\cite{Gasser:1984gg},
%and 
the lattice QCD simulations~\cite{Wilson:1974sk}.

In this letter, we will consider the NJL model %to study the phase
%diagram of quark matters. The NJL model is 
known as a low-energy
effective theory of QCD (for reviews, see,
\cite{Vogl:1991qt,Klevansky:1992qe,Hatsuda:1994pi,Buballa:2005rept}).
At low temperature, $T$, and chemical
potential, $\mu$, constituent quarks %have heavy constituent masses
are heavy due to the %underlying 
chiral symmetry spontaneous breaking %as observed in usual hadronic matters. 
while they are expected to be light at high $T$ and/or $\mu$ where
the chiral symmetry is getting restored. %restoration takes place. 
Thus the quark
system is closely related to the phenomenon of the chiral phase
transition.
%, and it is possible to study the phase structure of
%quark matters based on the argument of the chiral symmetry.
The NJL model actually predicts the chiral symmetry breaking at low
energy and its restoration at high energy.
% A lot of works are devoted to the
Many investigations of the phase diagram are based %through using
on the NJL and PNJL models (see, e.g.,~\cite{Wolff:1985av,
Hatsuda:1985eb,
Karsch:1986hm, Klimenko:1987gi, Asakawa:1989bq, Rosenstein:1990nm, 
Hands:1992ck, Zhuang:1994dw, Inagaki:1994ec, Berges:1998rc} 
and~\cite{Fukushima:2003fm,%
Ratti:2005jh, Roessner:2006xn, Fukushima:2008wg, Hell:2008cc}).

Since the NJL model is not renormalizable, the model predictions
inevitably depend on a regularization procedure applied. The most
frequently used method is probably the three-momentum %sharp 
cutoff
regularization %, cutoff regularization for short,
which introduces
the cutoff scale $\Lambda$. %in the three dimensional momentum
%space. 
%Although 
The model %with
in the cutoff scheme %successfully describes
%a hadronic system at low energy, it 
may miss an important
contribution when the quark density becomes comparable to the
cutoff scale.
%. Then we shall
%That is why we employ %the
%an alternative method,
%the dimensional regularization (DR), in this paper.
%%%%%%
There is an alternative method,
the dimensional regularization (DR), to avoid the issue 
\cite{Fujihara:2008ae}.
%%%%%%
In the
DR, divergences coming from fermion loop integrals are regularized
by lowering the dimension of the integration through an analytic
continuation in the dimension variable. Using various regularization ways is
interesting, because we believe that the regularization scheme is a dynamical
part of the NJL model, it is %procedure
related to the size and shape of the effective quark interaction
as discussed in~\cite{Inagaki:2011uj}.
It was found that the model with the DR nicely %predicts
describes quark systems
at low energy, such characteristics as the phase structure and meson properties%
~\cite{Fujihara:2008ae,Inagaki:2011uj,Inagaki:2007dq,Inagaki:2010nb}.

%The subject 
We shall study in this article %is
 the phase diagram %of
in the three flavor NJL model with the DR. It is interesting because
the recent work by the present authors~\cite{Inagaki:2011uj}
indicates that the phase structure, especially the order of the
transition, may %alter
differ drastically from the one in the cutoff
regularization. 
%Then our goal here is to draw and compare the
%phase diagrams of the NJL model with the DR and cutoff methods.

The structure of this letter is following: In Sec.~\ref{njl_model},
the three flavor NJL model and its parameters are presented.
Sec.~\ref{sec_critical} is devoted to the explanation on the
procedure of drawing the phase diagram. We then display the
resulting phase diagram of the model in Sec.~\ref{sec_pd}.
The concluding remarks are given in Sec.~\ref{conclusion}.

%%%%%%%%%%%%%%%%%%%%%%%%%%%%%%%%%%%%%%%%%%%%%%%%%%%%%%%%%%%%%%%%%%%%%%
%%%%%%%%%%%%%%%%%%%%%%%%%%%%%%%%%%%%%%%%%%%%%%%%%%%%%%%%%%%%%%%%%%%%%%
%    	Sec. 2 NJL MODEL WITH                                       %
%%%%%%%%%%%%%%%%%%%%%%%%%%%%%%%%%%%%%%%%%%%%%%%%%%%%%%%%%%%%%%%%%%%%%%
%%%%%%%%%%%%%%%%%%%%%%%%%%%%%%%%%%%%%%%%%%%%%%%%%%%%%%%%%%%%%%%%%%%%%%
\section{Three flavor NJL model}
\label{njl_model}
%%%%%%%%%%%%%%%%%%%%%%%%%%%%%%%%%%%%%%%%%%%%%%%%%%%%%%%%%%%%%%%%%%%%%%
\subsection{The model}
The Lagrangian of the three flavor model is %written by
\begin{eqnarray}
  \!\!\!\!\!\! \mathcal{L}_{\mathrm{NJL}}
  \!\!\!\! &=& \!\!\! \sum_{i,j} \overline{q}_i\left( i \partial\!\!\!/
       - \hat{m}\right)_{ij}q_j + \mathcal{L}_4 + \mathcal{L}_6 , 
 \label{LNJL} \\
  \mathcal{L}_4
  \!\!\!\! &=& \!\!\!  G \sum_{a=0}^8 \biggl[
      \Bigl( \sum_{i,j} \overline{q}_i\lambda_a q_j\Bigr)^2
     + \Bigl( \sum_{i,j}\overline{q}_i\,i \gamma_5 \lambda_a q_j \Bigr)^2
     \biggr] ,
  \label{L_4} \\
  \mathcal{L}_6
  \!\!\!\! &=& \!\!\! -K \left[ \det\overline{q}_i (1-\gamma_5) q_j 
     +{\rm h.c.\ } \right].
  \label{L_6}
\end{eqnarray}
where $\hat{m}_{ij}$ represents the diagonal mass matrix
${\rm diag}(m_u,m_d,m_s)$ with flavor indices $i,j$. $G$ and $K$ are
the four- and six-fermion couplings, $\lambda_a$ are the Gell-Mann
matrices in flavor space with $\lambda_0=\sqrt{2/3}\cdot {\bf 1}$.
The determinant in ${\mathcal L}_6$ runs over flavor space, so this
leads to the six-point interaction known as Kobayashi-Maskawa
't Hooft (KMT) term \cite{Kobayashi:1970ji}.

The vacuum of the model is determined by the minimum of the
thermodynamic potential $\Omega=-\ln Z/(\beta V)$ with the partition
function $Z$, the inverse temperature $\beta=1/T$, and the volume
of the system $V$. Applying the mean-field approximation, we can
calculate the potential $\Omega$ in the imaginary time formalism,
\begin{eqnarray}
  \Omega  \!\!\!\! &=& \!\!\! \Omega_v + \Omega_0 + \Omega_T ,\\
  \Omega_v \!\!\!\! &=& \!\!\!
   2G(\phi_u^2+\phi_d^2+\phi_s^2) -4K \phi_u \phi_d \phi_s\,,
   \label{omegav}\\
  \Omega_0 \!\!\!\! &=& \!\!\!
   -\frac{2^{D/2}N_c }{2}
    \int \!\! \frac{d^{D-1}p}{(2\pi)^{D-1}}
             \bigl[ E_u + E_d + E_s \bigr], \label{omega0}\\
  \Omega_T \!\!\!\! &=& \!\!\!
   -\frac{2^{D/2}N_c }{2}
   T \int \!\! \frac{d^{D-1}p}{(2\pi)^{D-1}}
             \sum_{i,\,\pm} \ln \Bigl[1 + e^{-\beta E_i^{\pm}} \Bigr].
  \label{omegat}
\end{eqnarray}
Here $\Omega_v$ corresponds to the vacuum contribution by the chiral
condensates, $\Omega_0$ and
$\Omega_T$ %indicate
denote the temperature independent and dependent
contributions, $\phi_i(\equiv \langle \bar{i}i \rangle)$ %expresses
is 
the chiral condensate for each quark which is the order parameter of
the model, $N_c(=3)$ is the number of colors. $D$ denotes dimensions
in the fermion loop integral, %which has an important role in the
%dimensional regularization scheme. 
$E_i =(p^2 + m_i^{*\,2})^{1/2}$
is the energy of the quasi-particle with the constituent quark mass
$m_i^*$, $E_i^{\pm}=E_i \pm \mu$ with a quark chemical
potential $\mu\,(=\mu_u=\mu_d=\mu_s$).

The fermion loop integral in Eq.(\ref{omega0}) %badly 
diverges, %and 
therefore we will perform the analytic continuation in $D$ to regularize
% the divergence
it by decreasing the dimension $D$ as discussed in%
~\cite{Inagaki:2010nb,Inagaki:2011uj}. %On the other hand 
In the
cutoff %way
scheme, the divergent contribution is dropped by introducing
the momentum cutoff $\Lambda$. To be more precise, the
regularization in the DR and cutoff schemes are performed by
the following %manipulations
replacements
\begin{eqnarray}
  \int \!\!\frac{d^{D-1}p}{(2\pi)^{D-1}} 
  \!\!\!\! &\rightarrow& \!\!\!\!
  \frac{2\,(4\pi)^{-(D-1)/2}}{\Gamma[(D-1)/2]} M_0^{4-D} \!\!\!
    \int_0^{\infty} \!\!dp \,p^{D-2} ,\\
  \int \!\!\frac{d^{D-1}p}{(2\pi)^{D-1}} 
  \!\!\!\! &\rightarrow& \!\!\!\!
  \frac{1}{2\pi^2} \int_0^{\Lambda} \!\!dp \,p^{2},
\end{eqnarray}
%%%%%%%%%%%%%%%%%%%%%%%%%%%%%%%%%%%%%%%%%%%%%%%%%%%%%%%%
where $M_0$ is the renormalization scale which
%plays the role to adjust the mass dimension of
is needed to render physical quantities %so that they have the
correct mass dimensions.

As mentioned in the introduction, the constituent quark mass 
\begin{equation}
m_i^* = m_i -4G \phi_i + 2K \phi_j \phi_k \,,\,\, (i\neq j \neq k)
\label{gap_eq}
\end{equation}
%%%%%%
is closely related to the chiral symmetry breaking,
namely to the value of $\phi_i$.
%%%%%%
The self-consistent gap equations~(\ref{gap_eq}) are obtained as the
condition for the thermodynamic
potential to be at the extremum, $\partial \Omega / \partial \phi_i=0$.
Eqs~(\ref{gap_eq}) explicitly show that the difference between
constituent and current quark masses is due to
the underlying chiral symmetry breaking.

%%%%%%%%%%%%%%%%%%%%%%%%%%%%%%%%%%%%%%%%%%%%%%%%%%%%%%%%%%%%%%%%%%%%%%
\subsection{Model parameters}
%%%%%%%%%%%%%%%%%%%%%%%%%%%%%%%%%%%%%%%%%%%%%%%%%%%%%%%%%%%%%%%%%%%%%%
The NJL model with the dimensional regularization has 7 free parameters:
current quark mass $m_u$, $m_d$, $m_s$, the four- and six-point couplings
$G$, $K$, the dimension $D$, and the renormalization scale $M_0$.

We consider, for simplicity, the isospin symmetric case, $m_d=m_u$, and
set several values for $m_u(=3,\,4,\,5,\,5.5,\,6{\rm MeV})$.
We then fix the remaining parameters by choosing 5 physical quantities
among listed below:
\begin{eqnarray}
  \begin{array}{cc}
    m_{\pi}=138{\rm MeV}, & f_{\pi}=92 {\rm MeV}, \\ 
    m_{\mathrm K}=495 {\rm MeV}, & m_{\eta^{\prime}}=958 {\rm MeV},\\
    m_{\eta}=548{\rm MeV}, & \chi^{1/4}=170{\rm MeV}.
  \end{array}
\label{input}
\end{eqnarray}
Following \cite{Inagaki:2011uj}, we name the parameter sets as
Case $\chi$ and $m_{\eta}$ depending on which quantities are selected.
The Case $\chi$ ($m_{\eta}$) is fitted by
$\{ m_\pi,f_{\pi},m_{\mathrm K}, m_{\eta^{\prime}}, \chi \,(m_{\eta}) \}$.
The parameter setting was performed in \cite{Inagaki:2010nb}, and we
shall employ three parameter sets, Case $m_{\eta}^{\rm LD}$, $m_{\eta}$ 
and $\chi$, which are shown in Tabs. \ref{para_etaLD}, \ref{para_eta}
and \ref{para_chi170}.
%--- Table ---%
\begin{table}[!h]
\begin{center}
 \caption{Case $m_{\eta}^{\rm LD}$.}
 \label{para_etaLD}
 %\begin{ruledtabular}
 \begin{tabular}{lccccc}
\hline
$m_u$ & $m_s$ & $G$ & $K$ & $M_0$ & $D$ \\
\hline %
$3.0$ & $84.9$ & $-0.0195$ & $9.02 \times 10^{-7}$ & $118$ & $2.29$ \\
%$4.0$ & $118$ & $-0.0174$ & $9.17 \times 10^{-7}$ & $113$ & $2.38$ \\
%$5.0$ & $156$ & $-0.0162$ & $9.49 \times 10^{-7}$ & $108$ & $2.47$
\hline
\end{tabular}
%\end{ruledtabular}
%\end{table}
%%%%%%%
%\begin{table}
 \caption{Case $m_{\eta}$.}
 \label{para_eta}
% \begin{ruledtabular}
 \begin{tabular}{lccccc}
\hline
$m_u$ & $m_s$ & $G$ & $K$ & $M_0$ & $D$ \\
\hline %
$3.0$ & $79.0$ & $-0.0130$ & $2.29 \times 10^{-7}$ & $107$ & $2.37$ \\
$4.0$ & $106$ & $-0.00748$ & $8.26 \times 10^{-8}$ & $92.0$ & $2.52$ \\
$5.0$ & $134$ & $-0.00357$ & $1.99 \times 10^{-8}$ & $73.2$ & $2.69$ \\
$5.5$ & $147$ & $-0.00231$ & $8.40 \times 10^{-9}$ & $62.4$ & $2.77$ \\
$6.0$ & $162$ & $-0.00142$ & $3.23 \times 10^{-9}$ & $50.9$ & $2.87$ \\
\hline 
 \end{tabular}
% \end{ruledtabular}
%\end{table}
%%%%%%%
%\begin{table}
 \caption{Case $\chi$.}
 \label{para_chi170}
% \begin{ruledtabular}
 \begin{tabular}{lccccc}
 \hline 
$m_u$ & $m_s$ & $G$ & $K$ & $M_0$ & $D$ \\
\hline %
$3.0$ & $77.1$ & $-0.0168$ & $2.23 \times 10^{-7}$ & $120$ & $2.28$ \\
$4.0$ & $106$ & $-0.0143$ & $2.11 \times 10^{-7}$ & $116$ & $2.36$ \\
$5.0$ & $134$ & $-0.0119$ & $1.80 \times 10^{-7}$ & $112$ & $2.43$ \\
$5.5$ & $150$ & $-0.0109$ & $1.62 \times 10^{-7}$ & $110$ & $2.47$ \\
$6.0$ & $166$ & $-0.00992$ & $1.48 \times 10^{-7}$ & $109$ & $2.50$ \\
\hline 
 \end{tabular}
% \end{ruledtabular}
\end{center}
\end{table}
%--- Table ---%
Note that the Case $m_{\eta}$ has two parameter sets for $m_u=3$MeV;
to distinguish between them we use the superscript ${\rm LD}$
(lower dimension).

For the sake of comparison we also align the parameters of the cutoff case
in Tab.~\ref{para_cut}.
%%%%%%
In the cutoff case, we fix 4 parameters,
$m_s, G, K$ and $\Lambda$ with $\{m_\pi, f_\pi, m_{\mathrm K}, m_{\eta'}\}$.
Unfortunately, there is no solution to simultaneously reproduce the 
above listed quantities for $m_u \gtrsim 5.87$MeV.
%%%%%%
%---- table-----%
\begin{table}[!h]
\begin{center}
 \caption{Case Cutoff.}
 \label{para_cut}
% \begin{ruledtabular}
{\tabcolsep = 4mm
\begin{tabular}{lcccc}
\hline 
$m_u$  & $m_s$  & $G\Lambda^2$ & $K\Lambda^5$ & $\Lambda$ \\
\hline %
$3.0$  & $89.5$ & $1.55$ & $8.34$ & $960$ \\
$4.0$  & $110$  & $1.60$ & $8.38$ & $797$ \\
$5.0$  & $128$  & $1.71$ & $8.77$ & $682$ \\
$5.5$  & $136$  & $1.81$ & $9.17$ & $630$ \\
$5.87$ & $139$  & $2.09$ & $10.1$ & $580$ \\
\hline 
\end{tabular}
}
% \end{ruledtabular}
\end{center}
\end{table}
%---- table-----%

%%%%%%%%%%%%%%%%%%%%%%%%%%%%%%%%%%%%%%%%%%%%%%%%%%%%%%%%%%%%%%%%%%%%%%
%%%%%%%%%%%%%%%%%%%%%%%%%%%%%%%%%%%%%%%%%%%%%%%%%%%%%%%%%%%%%%%%%%%%%%
%    	Sec. 3 Critical Behavior                                    %
%%%%%%%%%%%%%%%%%%%%%%%%%%%%%%%%%%%%%%%%%%%%%%%%%%%%%%%%%%%%%%%%%%%%%%
%%%%%%%%%%%%%%%%%%%%%%%%%%%%%%%%%%%%%%%%%%%%%%%%%%%%%%%%%%%%%%%%%%%%%%
\section{Critical behavior}
\label{sec_critical}

In this section we explain how to draw the phase diagram of
the model through the analysis %based on
of the thermodynamical
potential and the gap equations.

A critical temperature $T_c$ or chemical potential $\mu_c$ are given 
by the maxima of
\begin{equation}
  \frac{\partial \phi_u}{\partial t},
  \quad (t=T \,\, {\rm or} \,\,\, \mu) .
  \label{dPdT}
\end{equation}
In fact we apply $t=T\,(\mu)$ for low $\mu$ ($T$) in
crossover region. The above quantity becomes infinite at
$T_c\,(\mu_c$) when the transition is of the first order. In this case we
determine the transition boundary by the point where the
discontinuous change  of the chiral condensate $\phi_u$ occurs by
directly searching the minimum of the thermodynamic potential.
It is obvious that this procedure is consistent with the
criterion of Eq.~(\ref{dPdT}), because a divergent point coincides
the maximum point.

%%%%%%%%%%%%%%%%%%%%%%%%%%%%%%%%%%%%%%%%%%%%%%%%%%%%%%%%%%%%%%%%%%%%%
\subsection{Thermodynamic potential}
\begin{figure}[!h]
  \begin{center}
    \includegraphics[width=3.0in,keepaspectratio]{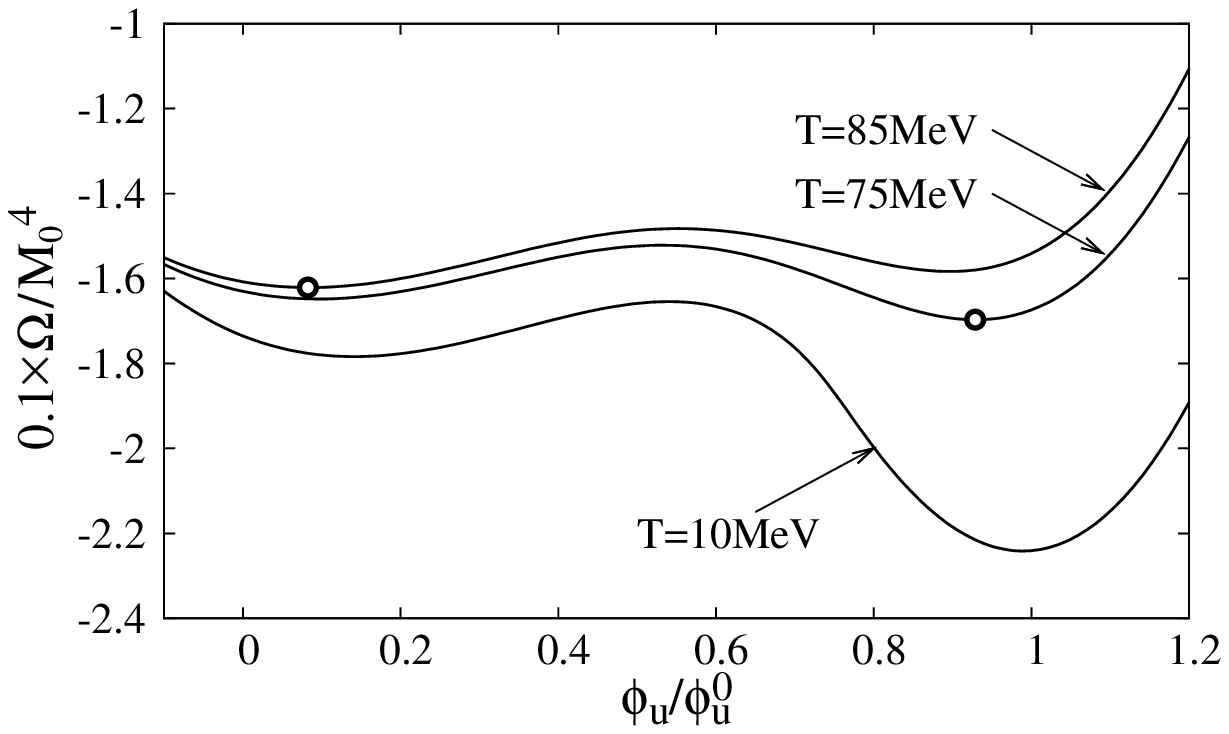}
    \includegraphics[width=3.0in,keepaspectratio]{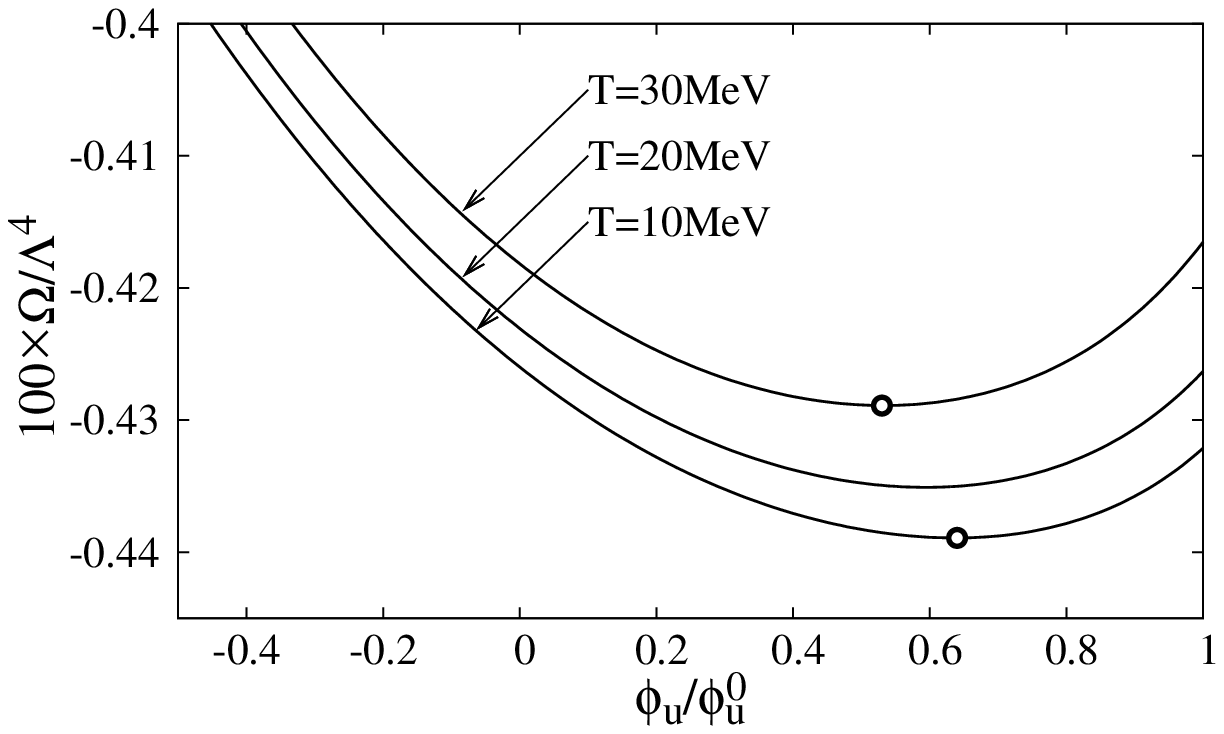}
  \end{center}
  \vspace{-0.3cm}
  \caption{Left panel: $0.1\cdot\Omega/M_0^4$ along the line
           $\phi_s=0.36\phi_u+0.83\phi_u^0$ in the Case $m_{\eta}$
           with $m_u=4$MeV for $T=10$, $75$, $85$MeV and $\mu=480$MeV.
           Right panel: $100 \cdot \Omega/\Lambda^4$ along the
           line $\phi_s=0.103\phi_u+1.43\phi_u^0$ in the Case Cutoff
           with $m_u=4$MeV for $T=10$, $20$, $30$MeV and $\mu=290$MeV.
           %Here $\Omega=\Omega(\phi_u,\phi_s)-\Omega(0,0)$
           %and $\phi_u^0$ denotes the chiral condensate
           %at $T,\mu=0$ for each case. 
           The circles indicate the global minima.}
  \label{thermo}
\end{figure}

%%%%%%%%%%%%%%%%%%%%%%%%%%%%%%%%%%%%%%%%%%%%%%%%%%%%%%%%%%%%%%%%%%%%%
To see the tendency of the phase transition, we show
the behavior of $\Omega(=\Omega(\phi_u,\phi_s)-\Omega(0,0))$ for the
Case $m_{\eta}$ and Cutoff with $m_u=4$MeV near the transition boundary
in Fig.~\ref{thermo}. 
%---- figure 1 -----%
%---- figure -----%
The curves are plotted %according to 
along the line
$\phi_s=0.36\phi_u+0.83\phi_u^0$ for $T=10$, $75$ and $85$MeV with
$\mu=480$MeV in the left panel, and %to
along the line
$\phi_s=0.103\phi_u+1.43\phi_u^0$ for $T=10$, $20$, $30$MeV with
$\mu=290$MeV in the right panel. These lines are chosen so as to show
the global minima for lower $T=75(10)$MeV and higher $T=85(30)$MeV,
which are %displayed 
indicated by the circles, near
the transition temperature $T_c\simeq80(20)$MeV. $\phi_u^0$ denotes
the chiral condensate $\phi_u$ at $T,\mu=0$ for each case.

%Here we observe 
The striking difference is observed between these figures. 
There exists a bump between two stable minima in the DR case, which
means that the transition is of the first order between $T=75$ and $85$MeV.
On the other hand %in the lower panel, 
the cutoff case (right panel) %shows 
produces rather
monotonous curves %and there appears
with no bump, which leads to a smooth
crossover. The difference stems from the fact that the ratio of the
thermal contribution ($\mu$ dependence)
$\Omega_T/(\Omega_v + \Omega_0)$ in the DR case is larger than that
in the cutoff case at low $T$. Thus we confirm the stronger tendency
of the first order phase transition in the DR scheme.

%The chiral condensates are determined by scanning the minimum of the
%thermodynamic potential. Therefore
%it will be the most direct approach to search the minimum of $\Omega$.
%At low $T$ and high $\mu$, this method becomes
%important because the potential has several local minima and which
%complicates the analyses by other ways.

%%%%%%%%%%%%%%%%%%%%%%%%%%%%%%%%%%%%%%%%%%%%%%%%%%%%%%%%%%%%%%%%%%%%%%
\subsection{$\partial \phi_u / \partial T$}
\begin{figure}
  \begin{center}
    \includegraphics[width=3.0in,keepaspectratio]{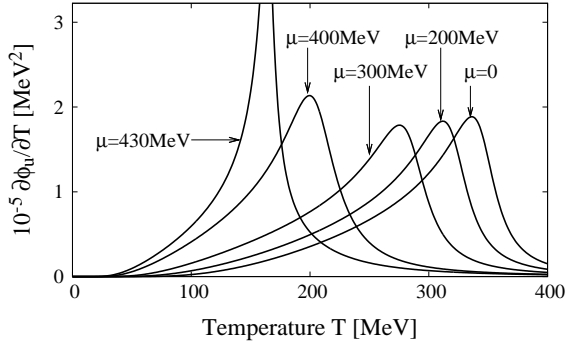}
  \end{center}
  \vspace{-0.3cm}
  \caption{$\partial \phi_u / \partial T$ in the Case $m_{\eta}$
           with $m_u=4$MeV.}
  \label{fig_dPdT}
\end{figure}

%%%%%%%%%%%%%%%%%%%%%%%%%%%%%%%%%%%%%%%%%%%%%%%%%%%%%%%%%%%%%%%%%%%%%%
In the crossover region, it is technically easier to analyze 
Eq.~(\ref{dPdT}) through solving the gap equations because $\phi_u$
changes continuously with respect to $T\,(\mu)$. We show the numerical results in Fig.~\ref{fig_dPdT}.
%---- figure 1 -----%
%---- figure -----%
One sees that the maximum point moves toward lower $T$ with
increasing $\mu$, and the peak becomes large at high $\mu$. The peak
actually diverges when $T$ and $\mu$ coincide with the critical point
$(T_{\rm CP},\mu_{\rm CP})$. Below $T_{\rm CP}$, the transition becomes
of the first order, and the analysis by Eq.~(\ref{dPdT}) is no longer
practically useful for the determination of the transition boundary
as mentioned above.

%%%%%%%%%%%%%%%%%%%%%%%%%%%%%%%%%%%%%%%%%%%%%%%%%%%%%%%%%%%%%%%%%%%%%%
%%%%%%%%%%%%%%%%%%%%%%%%%%%%%%%%%%%%%%%%%%%%%%%%%%%%%%%%%%%%%%%%%%%%%%
%    	Sec. 4 Phase diagram                                        %
%%%%%%%%%%%%%%%%%%%%%%%%%%%%%%%%%%%%%%%%%%%%%%%%%%%%%%%%%%%%%%%%%%%%%%
%%%%%%%%%%%%%%%%%%%%%%%%%%%%%%%%%%%%%%%%%%%%%%%%%%%%%%%%%%%%%%%%%%%%%%
\section{Phase diagram}
\label{sec_pd}
%%%%%%%%%%%%%%%%%%%%%%%%%%%%%%%%%%%%%%%%%%%%%%%%%%%%%%%%%%%%%%%%%%%%%%
We are now ready to discuss the phase structure of the NJL model
with the DR.

\subsection{Transition on $\phi_u$}

\begin{figure}
  \begin{center}
    \includegraphics[width=3.0in,keepaspectratio]{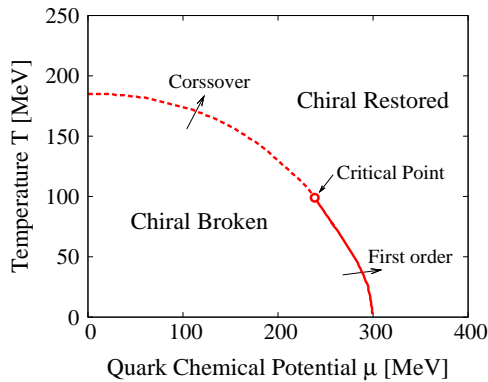}
  \end{center}
  \vspace{-0.3cm}
  \caption{Phase diagram in the Case $m_{\eta}^{\rm LD}$ with
           $m_u=3$MeV.
           The solid (dashed) line represents the first order
           (crossover) transition. The circle indicates the
           critical point.}
  \label{pd_etaLD}
\end{figure}

Fig.~\ref{pd_etaLD} displays the typical structure of the phase diagram
in the model with the DR %drawn 
in the Case~$m_{\eta}^{\rm LD}$.
%---- figure -----%
%---- figure -----%
%One observes the
This is a reasonable picture %in which the
of a system %is 
in the chiral symmetry
broken phase at low $T$ and $\mu$, and in the chiral symmetry restored phase at
high $T$ and/or $\mu$. The solid (dashed) line represents the first order
(crossover) transition, and the circle indicates the critical
point. Note that the transition temperature, $T_c=184$MeV for $\mu=0$, is
comparable with the lattice QCD prediction, $150-200$MeV. The critical
point is located %on
at $(T_{\rm CP},\mu_{\rm CP})=(99{\rm MeV},239{\rm MeV})$,
and it is interesting to see that $T_{\rm CP}$ is close to %the
one
obtained in the PNJL model with the cutoff
regularization, $T_{\rm CP}=102$MeV, for frequently used
parameter set %according to
of~\cite{Hatsuda:1994pi}, whereas 
%the corresponding case leads 
$T_{\rm CP}=48$MeV in the NJL model~\cite%
{Fukushima:2008wg}.
%%%%%%
%%%%%%
% It is also noticed
Note that the obtained critical point is close to 
one obtained in a NJL type model with the smooth form factor 
\cite{Berges:1998rc}, $(T_{\rm CP}, \mu_{\rm CP})=(101{\rm MeV},211{\rm MeV})$,
and in the linear sigma model \cite{Scavenius:2000qd}, 
$(T_{\rm CP}, \mu_{\rm CP})=(99{\rm MeV},207{\rm MeV})$.
Below we %shall 
make more detailed comparison between the
DR and the cutoff %methods.
schemes.

\begin{figure}
  \begin{center}
    \includegraphics[width=3.0in,keepaspectratio]{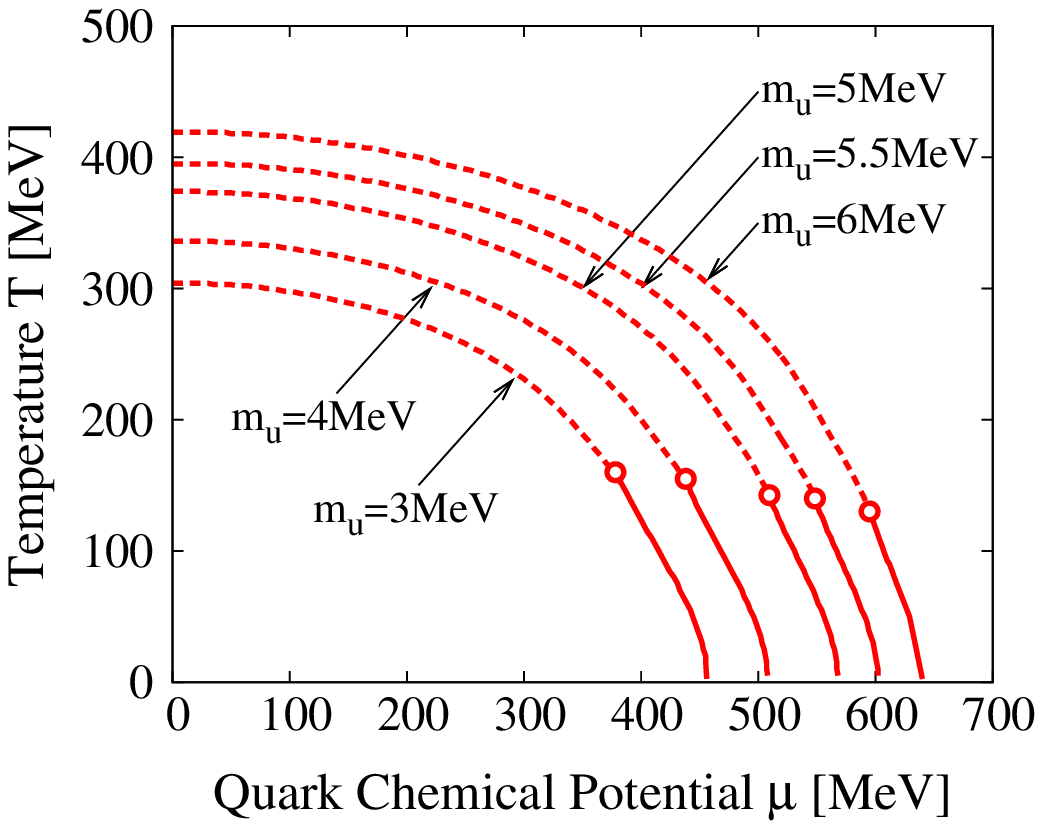}
    \includegraphics[width=3.0in,keepaspectratio]{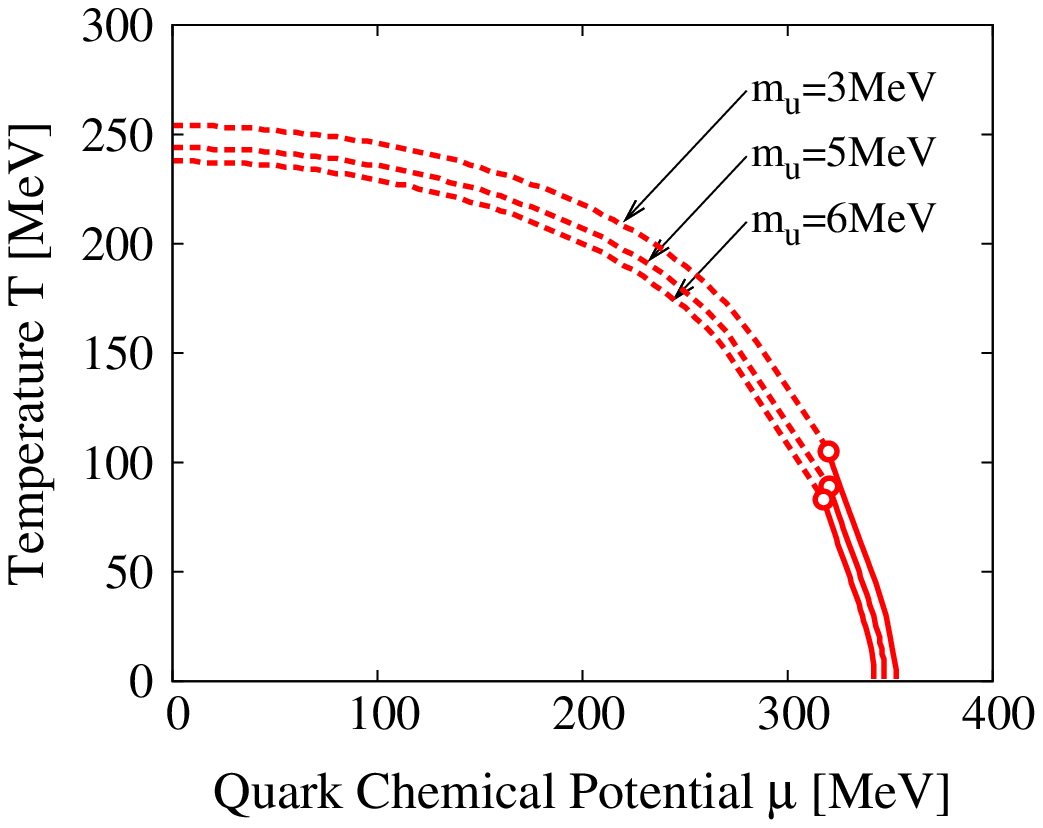}
  \end{center}
  \vspace{-0.3cm}
  \caption{Phase diagrams in the Case $m_{\eta}$ and $\chi$ are shown
           in the left and right panels. The solid (dashed) lines
           represent the first order (crossover) transition. The
           circles indicate the critical points.}
  \label{pd}
\end{figure}

Fig.~\ref{pd} shows the %resulting
 phase diagrams in the Cases~$m_{\eta}$
and $\chi$ for various $m_u$.
%---- figure -----%
%---- figure -----%
We note that in the Case $m_{\eta}$, the region of chiral symmetry broken phase
becomes smaller with choosing the smaller value of $m_u$. On the other
hand  the Case $\chi$ %shows the
produces similar curves for %each
different $m_u$. The
different behavior can be %understood
explained by the fact that the constituent
quark mass $m_u^*$ gets smaller with decreasing $m_u$ in the Case
$m_{\eta}$, while it %has similar value
almost does not change in the Case $\chi$ as
discussed in \cite{Inagaki:2010nb}.
%---- figure -----%
\begin{figure}
  \begin{center}
    \includegraphics[width=3.0in,keepaspectratio]{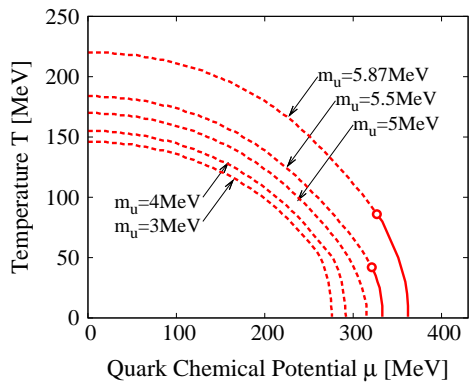}
  \end{center}
  \vspace{-0.3cm}
  \caption{Corresponding diagrams in the Case Cutoff.}
  \label{pd_cut}
\end{figure}
%---- figure -----%
In the cutoff case (Fig.~\ref{pd_cut}) %One sees that
 the region of the chiral symmetry broken state
shrinks when $m_u$ is lowered as observed in the Case $m_{\eta}$.
It is very interesting to note that the critical point disappears
below $m_u=5$MeV, where the transition is crossover for all $T$ and
$\mu$.

%Comparing the phase diagrams in the DR and cutoff regularization,
%we note a
A striking difference 
 between
the two regularizations is in that the critical point moves towards
higher temperature with decreasing $m_u$ in the DR, while it moves
to the opposite direction in the cutoff case. %This is indeed the sharp
%contrast seen between these two regularization schemes. 
The difference
may be understood by observing the value of the six-point coupling
$K$ which becomes larger (smaller) with decreasing $m_u$ in the DR
(cutoff) procedure, since the KMT term shown in Eq.(\ref{L_6}) tends
to drive the first order phase transition~\cite{Fukushima:2008wg}.

%%%%%%%%%%%%%%%%%%%%%%%%%%%%%%%%%%%%%%%%%%%%%%%%%%%%%%%%%%%%%%%%%%%%%%
\subsection{Partial transition on $\phi_s$}
\label{subsec_phi_s}
%%%%%%%%%%%%%%%%%%%%%%%%%%%%%%%%%%%%%%%%%%%%%%%%%%%%%%%%%%%%%%%%%%%%%%

As discussed in~\cite{Inagaki:2011uj}, the constituent quark masses
undergo two discontinuous changes at low $T$ in the DR scheme.
%---- figure -----%
\begin{figure}
  \begin{center}
    \includegraphics[width=3.0in,keepaspectratio]{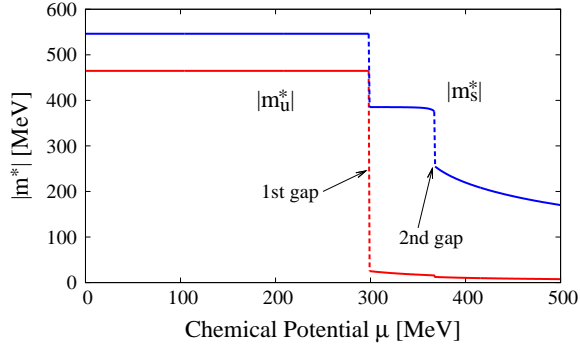}
  \end{center}
  \vspace{-0.3cm}
  \caption{Constituent quark mass $|m_u^*|$ and $|m_s^*|$ for
           $T=10$MeV in the Case $m_{\eta}^{\rm LD}$ with $m_u=3$MeV.}
  \label{aM_etaLD}
\end{figure}
%---- figure -----%
Fig.~\ref{aM_etaLD} displays the typical behavior of $|m_u^*|$ and
$|m_s^*|$ as functions of $\mu$ at low $T$ ($=10$MeV), plotted in
the Case $m_{\eta}^{\rm LD}$ with $m_u=3$MeV. One clearly
observes two gaps: one is located around $\mu_c^{(u)} \simeq 300$MeV and
the other is around $\mu_c^{(s)} \simeq 365$MeV. Here we call these
discontinuities as first and second gaps for lower and higher
chemical potential, respectively. The first gap comes from the
effect of the approximate ${\rm SU_L(2)\otimes SU_R(2)}$ % chiral
restoration and the second one %does 
comes from that of the partial
${\rm SU_L(3)\otimes SU_R(3)}$
restoration. Thus it may be interesting to study the phase
structure concerning %on
the second transition as well.

To draw the phase diagram on the second transition, we %will
set the criterion of the transition by using the following
quantity
\begin{equation}
  \frac{\partial \phi_s}{\partial t},
  \quad (t=T \,\, {\rm or} \,\,\, \mu).
  \label{dPsdT}
\end{equation}
%%%%%%
%{\bf ---Please make more clear this paragraph---}\\
%%%%%%
%The basic idea is essentially the same with the criterion by
%$\partial \phi_u/\partial t$. 
Then below $\mu_{\rm CP}$, namely in
the crossover region, the above quantity has only one maximum,
which determines the crossover transition on $\phi_s$. While
above $\mu_{\rm CP}$ the quantity
$\partial \phi_s / \partial \mu$ shows non-trivial behavior; it becomes
infinite at $\mu_c^{(u)}$, and has second maximum at $\mu_c^{(s)}$.
So $\partial \phi_s / \partial \mu$ has typical two maxima at
$\mu_c^{(u)}$ and $\mu_c^{(s)}$ below $T_{\rm CP}$ as %read 
seen in
Fig.~\ref{aM_etaLD}.  Here we %will 
call the transition point %concerning on 
corresponding to the second maximum, $\mu_c^{(s)}$,
%as 
``the second phase boundary". %here. 
To %clearly 
distinguish between the two
phase transitions, we call the transition line on $\phi_u$ discussed 
in the previous subsection %as 
``the first phase boundary". 
%%%%%%

%The transition on the second {\bf phase boundary} is defined at the point 
%where Eq.(\ref{dPsdT}) has the maxima except for the case $\phi_u$
%and $\phi_s$ show the first gap. It is obvious that the above
%definition is %the
%a straightforward extension of the $\phi_u$ case.

\begin{figure}
  \begin{center}
    \includegraphics[width=3.0in,keepaspectratio]{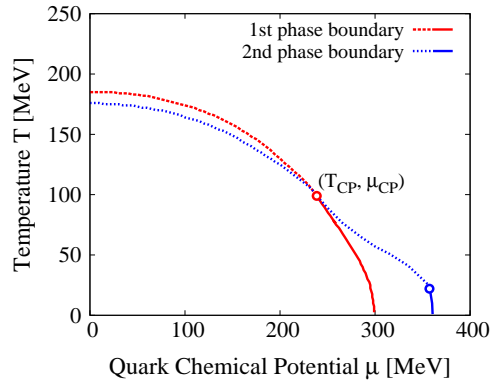}
  \end{center}
  \vspace{-0.3cm}
  \caption{The 1st and 2nd  phase boundaries in the Case
           $m_{\eta}^{\rm LD}$ with $m_u=3$MeV. The red dashed and
           blue dotted lines represent the crossover transition on
           $\phi_u$ and $\phi_s$. The red and blue solid lines
           indicate the transition on the first and second gap,
           respectively. The circles exhibit the critical points.}
  \label{pd_etaLD_2nd}
\end{figure}

%We show the resulting
In the phase diagram on the 1st and 2nd phase boundaries %in
(Fig.~\ref{pd_etaLD_2nd}) 
%---- figure -----%
%---- figure -----%
the dashed and dotted lines represent the crossover transition on
$\phi_u$ and $\phi_s$, respectively. The solid line for lower (higher)
chemical potential indicates the discontinuous change on the first (second)
gap.  We see that the crossover line on $\phi_s$ is observed at a bit
lower temperature than that on $\phi_u$ for $\mu < \mu_{\rm CP}$. It
should be noticed that the critical curves on $\phi_u$ and $\phi_s$
intersect at the critical end point ($T_{\rm CP},\mu_{\rm CP}$) on
$\phi_u$. Because the value of $\phi_s$ is
affected by $\phi_u$, as is clearly seen from Fig.~\ref{aM_etaLD}, %so
$\phi_s$ shows discontinuous change at the point where $\phi_u$ has
a gap. Then $\partial \phi_s/\partial t$ blows up and approaches to
infinity near the critical point where $\partial \phi_u/\partial t$
is divergent. Below $T_{\rm CP}$, $\partial \phi_s/ \partial \mu$ has
two maxima appearing at the first gap and higher chemical potential.
The first maximum coincides with the red solid line and the second
one is plotted by the blue line in Fig.~\ref{pd_etaLD_2nd}. The
transition on the second gap also has the critical point whose location
is exhibited by the blue circle at higher chemical potential.

We also studied the other Cases, $m_{\eta}$ and $\chi$, with various
$m_u$, and found that the qualitative behavior does not show remarkable
difference; the critical point on $\phi_s$ moves toward higher temperature
with decreasing $m_u$ as seen in the $\phi_u$ case. Therefore, we only
displayed the Case $m_{\eta}^{\rm LD}$ here.

%%%%%%%%%%%%%%%%%%%%%%%%%%%%%%%%%%%%%%%%%%%%%%%%%%%%%%%%%%%%%%%%%%%%%%
%%%%%%%%%%%%%%%%%%%%%%%%%%%%%%%%%%%%%%%%%%%%%%%%%%%%%%%%%%%%%%%%%%%%%%
%%      Sec. 5  Conclusion                                          %%
%%%%%%%%%%%%%%%%%%%%%%%%%%%%%%%%%%%%%%%%%%%%%%%%%%%%%%%%%%%%%%%%%%%%%%
%%%%%%%%%%%%%%%%%%%%%%%%%%%%%%%%%%%%%%%%%%%%%%%%%%%%%%%%%%%%%%%%%%%%%%
\section{Concluding remarks}
\label{conclusion}
%%%%%%%%%%%%%%%%%%%%%%%%%%%%%%%%%%%%%%%%%%%%%%%%%%%%%%%%%%%%%%%%%%%%%%
We studied the phase diagram of the NJL model with the DR and cutoff
regularization. % in this letter. Then
We found that the phase diagram
on the $T-\mu$ plane in the NJL model with the DR shows % the
a stronger tendency of the first order phase transition. The tendency is confirmed
by %observing
the shapes of the thermodynamic potential shown in Fig.~\ref{thermo}
where we find a bump in the DR, and rather monotonous %curve 
behaviour in the
cutoff case.

We have also studied the phase structure on the change of $\phi_s$
in Sec.~\ref{subsec_phi_s}., where we found that %both 
the approximate
${\rm SU_L(2)\otimes SU_R(2)}$ symmetry and the partial
${\rm SU_L(3)\otimes SU_R(3)}$ symmetry restore at %the 
a similar temperature for low chemical
potential, $\mu< \mu_{\rm CP}$. It may be
difficult to distinguish %both 
between the two lines experimentally, because the
transitions are smooth crossover at low chemical potential.

From the obtained phase diagrams, we % expect
conclude that the first order phase
transition persists for low $m_u$ in the model with the DR method.
%%%%%%
%%%%%%
%{\bf Please double check the next sentence!}
%%%%%%
The finding is consistent with the current 
symmetry analysis based consensus ~\cite{Pisarski:1983ms}
stating that the chiral phase transition %to be 
is of the first order % at low $T$.
 in the chiral limit, $m_{u,d,s} \rightarrow 0$.
%%%%%%
%%%%%%
This tendency may be understood by the following
reasoning. The loop contribution from %a lower momentum
the lower integration momenta is enhanced %for a lower 
by lowering dimension. It introduces %a
 non-locality in the model with the DR. %method. 
 The infrared behavior of the loop integral is important for %a
thermal corrections. It can rise the critical end point temperature, $T_{\rm CP}$. 
%of the critical end point.

%As a final remark, we pose the possibilities of further investigations.
Finally, because the parameter difference crucially affects the
location of the
critical point as confirmed in this article, %then
 we think it is
interesting to study the related issues, such as the case with the
chiral limit, and the $m_u,m_s$ dependence
on the order of the chiral transition in the context of the Columbia
plot~\cite{Brown:1990ev}.

%%%%%%%%%%%%%%%%%%%%%%%%%%%%%%%%%%%%%%%%%%%%%%%%%%%%%%%%%%%%%%%%%%%%%%
%%                         Acknowledgments                          %%
%%%%%%%%%%%%%%%%%%%%%%%%%%%%%%%%%%%%%%%%%%%%%%%%%%%%%%%%%%%%%%%%%%%%%%
\section*{Acknowledgments}
HK is supported by the grant NSC-99-2811-M-033-017 from National Science
Council (NSC) of Taiwan.

\end{document}